\documentclass[12pt]{article}
\usepackage{amsmath}
\usepackage{amssymb}

\input tcilatex
\QQQ{Language}{
British English
}

\begin{document}


\title{The Isotropy of Compact Universes}
\author{John D. Barrow$^1$ and Hideo Kodama$^2$ \\
$^1$DAMTP, Centre for Mathematical Sciences, \\
Wilberforce Rd., Cambridge CB3 0WA, UK\\
$^2$Yukawa Institute for Theoretical Physics, \\
Kyoto University, Kyoto 606-8502, Japan}
\maketitle

\begin{abstract}
We discuss the problem of the stability of the isotropy of the
universe in the space of ever-expanding spatially homogeneous
universes with a compact spatial topology. The anisotropic modes which
prevent isotropy being asymptotically stable in Bianchi-type $VII_h$
universes with non-compact topologies are excluded by topological
compactness. Bianchi type $V$ and type $VII_h$ universes with compact
topologies must be exactly isotropic. In the flat case we calculate
the dynamical degrees of freedom of Bianchi-type $I$ and $VII_0$
universes with compact 3-spaces and show that type $VII_0$ solutions
are more general than type $I$ solutions for systems with perfect
fluid, although the type $I$ models are more general than type $VII_0$
in the vacuum case. For particular topologies the 4-velocity of any
perfect fluid is required to be non-tilted. Various consequences for
the problems of the isotropy, homogeneity, and flatness of the
universe are discussed.
\end{abstract}


\section{Introduction}

The problem of providing a compelling explanation for the isotropy and
approximate flatness of the Universe has been the subject of extensive
analysis ever since the discovery of the temperature isotropy of the
microwave background radiation. Historically, a variety of approaches have
been taken to solving these problems. The first was to justify choosing
special initial conditions, at $t=0$ or $t=-\infty $, or 'almost' initial
conditions imposed at the Planck epoch $t\sim 10^{-43}s$ which marks the
threshold of quantum gravity. The second was to seek out general physical
processes which might transform a wide range of initial conditions into a
state that is similar to the presently observed universe after billions of
years of expansion (Misner 1967). This strategy was pursued first within the
context of the 'chaotic cosmology' programme during the period 1967-80. The
rules usually imposed were that classical dissipative stresses should obey
the strong energy condition of general relativity. Typical scenarios
considered involved the classical dissipation of anisotropy by collisional
or collisionless transport processes, and the depletion of irregularities by
the quantum particle production process (Doroshkevich, Zeldovich \& Novikov
1968, Stewart 1968, Collins \& Stewart,1971, Zeldovich and Starobinskii
1972, Barrow 1977, Barrow \& Matzner 1981).

One mathematically well-defined way of approaching the question of the
naturalness of the isotropy of the universe was to investigate the stability
of the isotropic Friedmann universes with respects to the set of all
anisotropic solutions of a given gravitation theory. This was first done by
Collins and Hawking (1973) in the context of spatially homogeneous general
relativistic cosmologies with zero cosmological constant by considering the
stability of non-collapsing isotropic universes in the space of Bianchi type
universes with the natural ${\mathbb R}^3$ topology. This reduces to the
study of the stability of particular solutions of ordinary differential
equations but is made non-trivial by the appearance of eigenvalues with
vanishing real part, and so the stability is generally determined at
non-linear order in the expansion around the isotropic universes in general.

The most general Bianchi universes containing the open Friedmann universe as
a special case are the Bianchi type $VII_h$ spaces. The open Friedmann
solution was shown by Collins and Hawking not to be asymptotically stable in
the space of all $VII_h$ initial data if matter obeyed the strong energy and
positive density conditions. A more detailed investigation (Barrow 1982,
Barrow \& Sonoda 1986) revealed that as time $t\rightarrow \infty $ isotropy
is stable (but not asymptotically stable) and identified the attractor as a
family of exact vacuum plane-wave spacetimes found by Lukash (1974). It
contains the isotropic Milne model as a special case.

The original result of Collins and Hawking is widely cited as showing that
isotropy was unstable (see, for example, Kolb and Turner 1990). However, the
asymptotic stability analysis of open universes is somewhat deceptive
because as $t\rightarrow \infty $ the imposition of the strong energy
condition ensures that these models approach vacuum (curvature-dominated)
solutions. Thus the asymptotic behaviour of the vacuum solutions ultimately
determines the asymptotic stability properties of the isotropic non-vacuum
solutions. However, our past cosmological history contains very little time
(if any at all) during which vacuum behaviour could have dominated the
Hubble expansion. Thus the asymptotic stability theorems are dictated by
conditions which have not existed in our past for any significant time
interval. They mainly tell us about the future. Thus, even if isotropy were
an asymptotically stable property of open sets of Bianchi $VII_h$ initial
data, it would not explain the isotropy of our universe, because it would
tell us only that isotropy is approached during the late curvature-dominated
phase of expansion at redshifts less than $z\sim 10.$

If the strong energy condition is violated then the conclusions will change.
Vacuum stresses of the sort that can drive inflation must violate the strong
energy condition in order to accelerate the mean scale factor of the
universe and isotropy can become asymptotically stable in accordance with
cosmic no hair theorems (Barrow 1982, Barrow \& Sonoda 1986). A particular
case of this result would be the asymptotic approach to the de Sitter metric
if a positive cosmological constant was admitted. Violation of the strong
energy condition is necessary but not sufficient for isotropization to occur
in this way.

If homogeneous initial conditions are confined to those in an open
neighbourhood of the spatially flat Friedmann universe then Collins
and Hawking (1973) went on to show that, in the space of Bianchi type
$VII_o$ initial data containing the flat Friedmann model as a special
case, isotropy is asymptotically stable if the matter content is
restricted to have zero pressure ('dust') to first order. The
condition for isotropization requires $\sigma /H\rightarrow 0$ as
$t\rightarrow \infty ,$ where $\sigma $ is the expansion shear and
shear and $H$ is the mean Hubble expansion rate.  Isotropy is not
approached during in radiation-dominated models of this Bianchi
type. Thus if there was some reason for the initial data to be of zero
curvature ('flat') it appeared that isotropy might be regarded as an
asymptotically stable cosmological feature. However, again, the
observational consequences are limited because there have been so many
more e-foldings of cosmic expansion during an era of radiation
domination (from $ 10^{-43}s$ to $10^{10}s$) than one dominated by
dust (from $10^{10}s$ to $ 10^{17}s$) during the history of the
universe (Barrow 1982; Barrow \& Sonoda, 1985, 1986). Also, as in the
open universe case, if the strong energy condition and zero pressure
condition restrictions are relaxed to permit the inclusion of stresses
with $\rho +3p<0$, where $\rho $ is the matter density and $p$ is the
pressure then isotropy becomes asymptotically stable for an open set
of initial data. This situation has been investigated in detail by
Wainwright et al (1999) and Nilsson et al (2000) who discriminate
between isotropization of the expansion shear, $\sigma$, and that of
the Weyl curvature anisotropy relative to the mean Hubble expansion
rate, $H$. Radiation and dust-dominated $VII_0$ universes exhibit
isotropic shear isotropization, $\sigma /H\rightarrow 0$ as
$t\rightarrow \infty $ but the Weyl curvature anisotropy, $W/H$ grows
in the radiation dominated era and approaches a non-zero constant
(which need not be small) in the dust era as $t\rightarrow \infty $
(Collins and Hawking did not require the Weyl curvature anisotropy to
tend to zero asymptotically in order for isotropization to occur in
type $VII_0$). This feature was also stressed by Doroshkevich et al
(1973) and Lukash (1974). Unlike the shear anisotropy, a large Weyl
anisotropy does not require a large temperature anisotropy in the
microwave background. The type $VII_h$ asymptote does have
$W/H\rightarrow 0$ as $t\rightarrow \infty $ because it is a
plane-wave spacetime for which all scalar curvature invariants are
zero.

These results were interpreted as showing that if the strong energy
condition holds (no inflation allowed) then generic spatially homogeneous
ever-expanding anisotropic universes do not become isotropic during their
expansion histories. A separate analysis of closed Bianchi type IX universes
with $S^{3\:}$topology shows that isotropy is unstable under the
similar conditions (Doroshkevich et al 1973). Of course, if the strong
energy condition is dropped then a finite period of inflationary expansion
in the past can reduce any initial anisotropy observed inside our horizon
below any given level by a pre-specified time and drive the expansion of any
open or not too closed universe very close to flatness for very long
intervals of time at late times without either isotropy or flatness being
approached as $t\rightarrow \infty $. If it is too closed it may collapse
before inflation can occur. Moreover, if further conditions of physical
reality are imposed, so that the anisotropy energies in long-wavelength
gravitational wave degrees of freedom at the Planck epoch do not
significantly exceed the Planck density then the size of anisotropies today
is constrained to be quite close to that observed in the microwave
background (Barrow 1995).

The basis of these results about the stability of isotropic universes when
the strong energy condition holds is the instability of isotropic universes
to perturbations by spatially homogeneous anisotropic modes. The most
general Bianchi type universes are characterized by $4$ constant parameters
on a Cauchy surface of constant time in vacuum and by $8$ parameters in the
presence of a perfect fluid (Ellis and MacCallum 1969). The most general
Bianchi types containing the flat, open, and closed Friedmann universes are
of Bianchi types $VII_0,VII_h,IX$, respectively and are specified by $3,4,4$
constant parameters in vacuum and $7,8,8$ parameters in the perfect fluid
case, respectively. For comparison, in the presence of a perfect fluid, the
flat Friedmann universe has no free parameters, and the open and closed
models have $1$ free parameter.

All these results assume that the topology of ever-expanding universes is
the natural ${\mathbb R}^3$ topology so that they have infinite spatial
volume if the 3-curvature is never positive. We shall show that the
conclusions change significantly if it is assumed that the topology of the
spatial sections of spatially homogeneous anisotropic universes of
non-positive curvature are \textit{spatially compact}.


\section{Spatially Compact Bianchi type universes}

There has been considerable interest in the possibility that the
Universe might possess compact space sections of non-positive
curvature (see the conference proceedings collection edited by
Starkman1998). These considerations have been motivated in part by the
'naturalness' of finite spatial sections in quantum cosmologies. There
is a long history of occasional investigations of the observational
consequences of compact flat universes with 3-torus topology
(Lachieze-Rey and Luminet 1995) and the possibility was even
considered by Friedmann (1924) to show that open universes were not
necessary spatially infinite. Studies of these compact cosmologies
have been dominated by studies of multiple imagery (Sokolov and
Shvartsman 1975) but there has been also some work which related the
observed homogeneity and isotropy of the universe to its spatial
topology. For example, the possibility of invoking a 'small' compact
open or flat universe to account for the large scale uniformity of the
universe has been proposed by Ellis and Schreiber (1986) following the
pioneering consideration of topological effects on local cosmological
structure by Ellis (1971). They have explored the possibility that the
cosmological horizon problem can be significantly ameliorated by
identifying points in such a way that observers can see all the way
round the universe after some given time. Although this construction
is easy to achieve in flat Friedmann universes, it is necessary to
explore its effects upon a fully anisotropic and inhomogeneous
universe. In a related study, Ellis and Tavakol (1994) have considered
the effects of geodesic mixing in compact open universes on the
propagation of microwave background photons after last scattering and
the topological constraints on anisotropy damping by this process were
considered by Reboucas et al (1998).

The rapid progress in observations of the cosmic microwave background
is also providing a possibility to observe directly the topology of
the universe. In particular, it has recently been recognized that the
changed appearance of the microwave background sky pattern offers a
sensitive probe of the spatial topology if the scale of periodicity is
sufficiently close to the particle horizon scale (Levin et al, 1997,
1999, Cornish, Spergel and Starkman 1998). In the flat Friedmann
universe there is no reason why these two scales should be
similar. However, open universes provide a physical curvature scale
which might be closely related to the overall periodicity scale. For
this reason there has been much recent interest in the observational
features of compact open Friedmann universes.  Their behaviour is
considerably more diverse than that of compact flat
universes. Geodesics display chaotic divergences on negatively curved
spaces and possible topologies are extremely complex, the eigenmode
problem is unsolved in general, and their specifications are only
partially understood.

Let us now consider an extension of the problem of 'why the universe is
isotropic?' to the situation of compact, spatially homogeneous, anisotropic
universes. We will consider the compactification of Bianchi type universes
with zero and negative curvature. As in the investigation made by Collins
and Hawking (1973) for the non-compact case, we first ask what are the most
general Bianchi types which contain the flat and open Friedmann universes as
particular cases. This problem has been studied by several authors (Ashtekar
and Samuel 1991, Fagundes 1985, 1992, Kodama 1998, Koike et al 1993,
Tanimoto et al 1997, 1997a).

If a group contains a subgroup $\Gamma $ which acts on a manifold $X$ as a
covering group so that $X/\Gamma $ becomes compact then the geometry of the
manifold is said to admit a compact quotient.
Thurston (1979) classified all maximal simply-connected three-dimensional
geometries which admit a compact quotient into one of eight possible cases.

The Bianchi classification of spatially homogeneous universes is
derived from that of the three-dimensional Lie groups that act freely
on a four-dimensional spacetime manifold as an isometry group with
spatial orbits. We say that a homogeneous manifold is simply
homogeneous if the group of isometries has a three-dimensional
subgroup that acts simply transitively on the manifold. The Bianchi
types are subdivided into two classes (Ellis and MacCallum, 1969):
Class A contains types $ I,II,VI_0,VII_0,VIII$ and $IX$ while Class B
contains types $IV,V,III,VI_h$ and $VII_h$. Except for types $IV$ and
$VI_h$, we can construct a spacetime manifold with compact space for
these models by suitable identifications of their spatially
homogeneous hypersurfaces. This operation will usually lower the
dimension of the isometry group so that the spatial hypersurfaces of
the resulting spacetime is no longer simply homogeneous. For example,
it is known that a homogeneous space section in any Class B model
cannot be simply homogeneous if it is compact. Compact quotients are
often not homogeneous globally at all, (i.e., the full symmetry group
does not act transitively). This happens even for the simple manifold
$T^3/{\mathbb Z}_2$ which is a quotient of the Euclidean space $E^3$
by a discrete group generated by two translations in the $x-y$ plane
and a combination of a translation along the $z$-axis and a rotation
by angle $\pi $ around the same axis. Therefore, in general, the
spacetime with compact space can only be locally homogeneous. The same
argument applies to isotropy as well.

Further new features appear in the compact cases. First, a non-trivial
compact topology may require the geometry and matter configurations to have
a higher symmetry than that of a simply transitive group when the data is
lifted to the covering spacetime. Such a situation arises when the
identification group $\Gamma $ cannot be contained in a simply transitive
group. This feature has an important consequence for the isotropization
problem, as we will see below. Second, new dynamical degrees of freedom,
called 'moduli parameters', appear. These moduli parameters describe
globally non-isometric deformations of the geometry which preserve the local
geometric structure. For some cosmological models, the total number of
dynamical degrees of freedom becomes much larger than that for the
non-compact case due to the existence of the moduli degrees of freedom.


\subsection{Open universes}

The introduction of compactness imposes a major constraint upon homogeneous
anisotropic universes with negative spatial curvature. The Bianchi types $V$
and $VII_h$ contain open Friedmann universes as special cases. By inspecting
the subgroup structure of each maximal symmetry group, one finds that these
groups must be subgroups of the maximal symmetry corresponding to the
Thurston type $H^3$ if they act simply transitively on constant-time spatial
sections $\tilde \Sigma $ of the covering spacetime. Hence, by Thurston's
theorem, if the universal covering $(\tilde M,\tilde g)$ of a locally
homogeneous spacetime $(M,g)$ with compact space has a symmetry group $G$
containing a simply transitive subgroup of Bianchi type $V$ or $VII_h$, then 
$G$ must be a subgroup of the isometry group of $H^3$, and so $(M,g)$ is
written as $(\tilde M,\tilde g)/\Gamma $ with some discrete group $\Gamma
\subset G$.

It is here, when determining minimal possible $G$, that the Class A or
B nature of the homogeneity group becomes important. For the Class B
Bianchi space sections, $\tilde \Sigma$, there exists a non-vanishing
vector field $v$ written as $q^{IJ}c_IX_J$ where $X_I$ is an invariant
basis with a structure constant $c^I{}_{JK}$, $c_I=c^J{}_{IJ}$ and
$q_{IJ}$ is the component of a homogeneous spatial metric with respect
to $X_I$. Since this vector has non-vanishing divergence, $\nabla
\cdot v=g^{IJ}c_Ic_J$, it cannot be invariant under the action of
$\Gamma $. For, if it were invariant, it would create a well defined
vector field $v^{\prime }$ with non-vanishing divergence on the
quotient $\Sigma =\tilde \Sigma /\Gamma $.  But this would lead to a
contradiction since

\[
0=\int_\Sigma d^3x\sqrt{g}\nabla \cdot v^{\prime }=\nabla \cdot v^{\prime
}\int_\Sigma d^3x\sqrt{g}{\neq }0. 
\]

However, one finds that if $G$ is smaller than the connected component of
the full isometry group of $H^3$, $G$ keeps $v$ invariant. Therefore the
covering spacetime must be invariant under a group isomorphic to the maximal
isometry group of $H^3$ and be spatially homogeneous and isotropic.

Furthermore there is no moduli freedom in the Class B case from the Mostow
rigidity theorem (Thurston 1979, 1982), which says that two compact
hyperbolic spaces are isometric up to a constant scaling if they are
homeomorphic. Thus compact universes of Bianchi types $V$ and $VII_h$ admit
only an overall change of volume scaling factor. They must therefore be
isotropic. Hence, we have the rather surprising result that compact open
universes of type $VII_h$ not only make isotropy an asymptotically stable
property of the initial data set but they permit no anisotropy to be present
at all. Recall that in the non-compact stability analyses of the open
Friedmann models the asymptotic behaviour was approach to a particular
2-parameter set of anisotropic type $VII_h$ plane-wave universes in which
the ratio of the shear to Hubble scalar is a constant (when this ratio is
zero the isotropic vacuum Milne universe is obtained). These anisotropic
solutions are not admitted when the space sections are compactified.
Compactification requires that Bianchi type initial data containing
isotropic universes must be exactly isotropic.


\subsection{Flat universes}

The flat Friedmann universe is a special case of Bianchi types $I$ and
$VII_0 $. In these Class A universes the introduction of
compactification does not restrict the symmetry group so strongly as
in the Class B cases, and anisotropy is permitted. However, the
spatial topology still gives some weak constraints on the minimal
symmetry of the universal covering as shown in Table
\ref{tbl:count:E3}. For example, when the spatial sections are
homeomorphic to $T^3$, $T^3/{\mathbb Z}_2$ or $T^3/{\mathbb Z}_2\times
{\mathbb Z}_2$, under an appropriate choice of coordinates, the
minimal symmetry of the covering data with the Bianchi I symmetry is
given by ${\mathbb R}^3{\tilde \times }D_2$, while for $T^3/{\mathbb
Z}_k$($k=3,4,6$) the minimal symmetry is ${\mathbb R}^3{\tilde \times }
O(2)$. Here, $D_2$ is the dihedral group $\{1,R_1(\pi ),R_2(\pi ),
R_3(\pi )\}$ consisting of rotations by angle $\pi $ around three
orthogonal coordinate axes; $O(2)$ is the group generated by rotations
around the $z$-axis and $D_2$, and $A{\tilde \times }B$ is a
semi-direct product of two groups $A$ and $B$. Hence, the
isotropization behavior of the former group is the same as that of the
generic open type I case, but the behavior for the latter group
coincides with that of axisymmetric type I model. However, isotropy is
an asymptotically stable property for spatially compact Bianchi I
models with perfect fluid satisfying the dominant energy
condition. The Bianchi type I geometry does not permit the presence of
non-comoving velocities.

Similarly, for $T^3$ and $T^3/{\mathbb Z}_2$, the minimal symmetry of
the covering data with type $VII_0$ symmetry coincides with the
$VII_0$ group itself, hence the isotropization behavior is determined
by that of generic open type $VII_0$ models with non-vanishing generic
fluid velocity. In contrast, for $T^3/{\mathbb Z}_2\times {\mathbb
Z}_2$ and $T^3/{\mathbb Z}_k(k=3,4,6)$, the minimal symmetry is
$VII_0{\tilde \times }D_2$, which requires that the space metric be
diagonal and the fluid 4-velocity thus be orthogonal to the spacelike
hypersurfaces of constant time. Since the isotropic type $VII_0$ model
with radiation is already unstable against perturbations with the
latter symmetry, (Wainwright et al 1999, Nilsson et al 2000)
compactification does not change the conclusion on the isotropization
of type $VII_0$ Bianchi models.

\begin{table}[tbp]
\begin{tabular}{llccccccc}
\hline\hline
\textbf{Symmetry} & \textbf{Space} & $Q$ & $P$ & $N_m$ & $N_f$ & $N$ & $N_s$
& $N_s$(vacuum) \\ \hline
&  &  &  &  &  &  &  &  \\ 
&  &  &  &  &  &  &  &  \\ 
${\mathbb R}^3{\tilde\times} D_2$ & ${\mathbb R}^3$ & 0 & 3 & 0 & 1 & 4 & 2
& 1 \\ 
& $T^3$ & 3 & 3 & 6 & 1 & 13 & 11 & 10 \\ 
& $T^3/{\mathbb Z}_2$ & 3 & 3 & 2 & 1 & 9 & 7 & 6 \\ 
& $T^3/{\mathbb Z}_2\times{\mathbb Z}_2$ & 3 & 3 & 0 & 1 & 7 & 5 & 4 \\ 
&  &  &  &  &  &  &  &  \\ 
VII$_0$ & ${\mathbb R}^3$ & 2 & 3 & 0 & 4 & 9 & 7 & (3) \\ 
& $T^3$ & 3 & 3 & 4 & 4 & 14 & 12 & (8) \\ 
& $T^3/{\mathbb Z}_2$ & 3 & 3 & 2 & 4 & 12 & 10 & (6) \\ 
&  &  &  &  &  &  &  &  \\ 
VII$_0{\tilde\times}{\mathbb Z}_2$ & ${\mathbb R}^3$ & 2 & 3 & 0 & 2 & 7 & 5
& (3) \\ 
& $T^3$ & 3 & 3 & 4 & 2 & 12 & 10 & (8) \\ 
& $T^3/{\mathbb Z}_2$ & 3 & 3 & 2 & 2 & 10 & 8 & (6) \\ 
&  &  &  &  &  &  &  &  \\ 
VII$_0{\tilde\times} D_2$ & ${\mathbb R}^3$ & 2 & 3 & 0 & 1 & 6 & 4 & 3 \\ 
& $T^3$ & 3 & 3 & 4 & 1 & 11 & 9 & 8 \\ 
& $T^3/{\mathbb Z}_2$ & 3 & 3 & 2 & 1 & 9 & 7 & 6 \\ 
& $T^3/{\mathbb Z}_2\times{\mathbb Z}_2$ & 3 & 3 & 1 & 1 & 8 & 6 & 5 \\ 
& $T^3/{\mathbb Z}_k(k=3,4,6)$ & 3 & 3 & 0 & 1 & 7 & 5 & 4 \\ 
&  &  &  &  &  &  &  &  \\ 
${\mathbb R}^3{\tilde\times} O(2)$ & ${\mathbb R}^3$ & 0 & 2 & 0 & 1 & 3 & 1
& 0 \\ 
& $T^3$ & 2 & 2 & 4 & 1 & 9 & 7 & 6 \\ 
& $T^3/{\mathbb Z}_2$(aligned) & 2 & 2 & 2 & 1 & 7 & 5 & 4 \\ 
& $T^3/{\mathbb Z}_2$(oblique) & 2 & 2 & 3 & 1 & 8 & 6 & 5 \\ 
& $T^3/{\mathbb Z}_2\times{\mathbb Z}_2$ & 2 & 2 & 1 & 1 & 6 & 4 & 3 \\ 
& $T^3/{\mathbb Z}_k(k=3,4,6)$ & 2 & 2 & 0 & 1 & 5 & 3 & 2 \\ 
&  &  &  &  &  &  &  &  \\ 
$ISO(3)$ & ${\mathbb R}^3$ & 0 & 1 & 0 & 1 & 2 & 0 & 0 \\ 
& $T^3$ & 1 & 1 & 5 & 1 & 8 & 6 & 5 \\ 
& $T^3/{\mathbb Z}_2$ & 1 & 1 & 3 & 1 & 6 & 4 & 3 \\ 
& $T^3/{\mathbb Z}_2\times{\mathbb Z}_2$ & 1 & 1 & 2 & 1 & 5 & 3 & 2 \\ 
& $T^3/{\mathbb Z}_k(k=3,4,6)$ & 1 & 1 & 1 & 1 & 4 & 2 & 1 \\ \hline\hline
\end{tabular}
\caption{The number of dynamical degrees of freedom for locally homogeneous
systems with perfect fluid on the spaces of Thurston type $E^3$. $N_s$
represents the dimension of the solution space. See the Appendix for the
details of the notation.}
\label{tbl:count:E3}
\end{table}


\subsection{Parameter Counting}

A parameter-counting classification of Bianchi type universes can be
performed in the compact case, as is explained in the Appendix in detail.
Some of the constant parameters represent the moduli freedom, but in general
it is difficult to distinguish clearly between the dynamical freedom and the
moduli freedom. The general result of Kodama (1998), Koike et al (1993), and
Tanimoto et al (1997, 1997a) is that the number of dynamical degrees of
freedom, i.e., the dimension of the solution space, becomes larger in the
compact cases than in the non-compact cases for the Class A Bianchi
universes. This is partly due to the appearance of the moduli degrees of
freedom and partly due to the decrease in the freedom of diffeomorphisms
connecting physically equivalent solutions. For example, in the case of
simple 3-torus, $T^3$, we can obtain the correct parameter count for the
vacuum case by the following naive argument. First, in order to specify the
lattice in the Euclidean 3-space so as to define the 3-torus in a
rotationally invariant way, we need 3 parameters to specify the lengths of 3
vectors generating the lattice and 3 parameters to specify the relative
direction angles of these vectors. Hence, adding their time derivatives, we
need 12 parameters. If we take into account the Hamiltonian constraint and
the time translation freedom, the total number reduces to 10.

\begin{table}[tbp]
\begin{tabular}{ccccc}
Bianchi Type & \multicolumn{2}{c}{Vacuum} & \multicolumn{2}{c}{Perfect fluid}
\\ 
& Non-compact & Compact & Non-compact & Compact \\ 
$I$ & 1 & 10 & 2 & 11 \\ 
$II$ & 2 & 6 & 5 & 9 \\ 
$VI_0$ & 3 & 4 & 7 & 8 \\ 
$VII_0$ & 3 & 8 & 7 & 12 \\ 
$IX$ & $-$ & 4 & $-$ & 8
\end{tabular}
\caption{Maximal degrees of freedom for spatially open and compact Bianchi
vacuum models and for those with perfect fluid.}
\label{tbl:count}
\end{table}

One significant feature of the parameter count in the spatially
compact open universes is the difference between the counts for vacuum
and perfect fluid models. If we separate the moduli and dynamical
freedoms in the way explained in the Appendix, the
diffeomorphism-invariant phase space can be, roughly speaking, written
as the product of the phase space of the diagonal homogeneous system
and the moduli space, at least for the vacuum Bianchi $I,II,VI_0$ and
$VII_0$ systems. We can show that the moduli parameters defined in
this sense become constants of motion. Hence, the dynamics are
essentially represented by the diagonal system with $3+3$
parameters. The Hamiltonian constraint should then also be imposed. In
this representation, the phase space of the non-compact case is
obtained by discarding the moduli freedom and taking into account the
additional equivalence relation for the diagonal system. For example,
in the type $I$ case, $3$ metric coefficients of the diagonal system
can be set to unity by a change of spatial coordinates at the initial
time, and we obtain a final count of $6-3=3$. The Hamiltonian
constraint and the temporal gauge fixing reduces it further to
$3-2=1$. In the compact case, we cannot do such a rescaling because it
changes the moduli parameters, and the count is $10$. From Table \ref
{tbl:count:E3}, we find that it is $8$ for the vacuum $VII_0$ system
on $T^3$. This implies that among the spatially compact \textit{vacuum
}Bianchi models, the type $I$ model is the most generic unlike in the
usual situation with non-compact topology, where it is the least
generic.

In contrast, when a perfect fluid is present, Bianchi type $VII_0$ universes
need not be diagonalizable because the momentum constraints simply relate
the non-diagonal components of the extrinsic curvature to the spatial
components $u_I$ of the fluid 4-velocity. Since the momentum constraints
require $u_I$ to vanish for type $I$ models, it turns out that the parameter
count for type $VII_0$ models is always larger than that for type $I$ models
in any given space topology. Therefore, the most general locally homogeneous 
\textit{perfect-fluid} spacetimes that include the flat isotropic model are
the Bianchi type $VII_0$ models in the spatially compact case just as they
are in the non-compact case.

In the table we have also included, for comparison, the parameter counts for
the compact Bianchi type $IX$ universes with $S^3$ topology which contain
the closed Friedmann universes as isotropic sub-cases. The vacuum $IX$
universes are $4$-parameter, while the perfect fluid type $IX$ universes are 
$8$-parameter.

Here note that it is easy to extend these parameter counts to systems with 
scalar fields. Since the homogeneity-group preserving diffeomorphisms act 
trivially on scalar fields  and scalar fields do not contribute to the 
momentum constraint in a locally homogeneous spacetime, the parameter 
count simply increases by two for each real component of the scalar fields 
irrespective of the Bianchi type and the other matter contents. Hence, the 
compact Bianchi I models are still more general than the compact Bianchi 
VII models if the models contain only scalar fields.

When counting 'parameters' in the compact cases it is important to remember
that the lengths and the angles of the identifications are time-dependent
variables specifying the compact spatial geometry $g_{ij}$ at each time.
Since the Einstein equation is second order in time, we must also specify
their time derivatives, $\dot{g}_{ij},$ at an initial time, and then their
future time development is completely determined. For the simple 3-torus
case, six combinations of these variables and their time derivatives become
the moduli parameters, which turn out to be constant in time because the
Hamiltonian constraint does not depend on them. The other six combinations
correspond to the diagonalized metric components and their time derivatives.
However, such a simple counting argument sometimes fails for other
topologies because the canonical structure becomes degenerate in the moduli
sector; that is, some moduli parameters do not have conjugate momenta, as
was shown in Kodama (1998). Since the existence and the uniqueness up to
diffeomorphisms of the generic initial value problem holds for the Einstein
equations, we can determine the number of independent solutions even in such
cases by calculating the dimension of the locally homogeneous sector of the
full diffeomorphism-invariant phase space. This is the approach adopted in
Kodama (1998), and differs from that adopted in the papers by Koike et al
(1993), and Tanimoto et al, (1997, 1997a) in which spacetime solutions are
classified. In the latter approach, knowledge on the explicit form of
spacetime solutions is required.

Finally, note that in contrast to the cases where space has ${\mathbb R}^3$
topology, it is meaningless to compare the parameter counts for spatially
compact Bianchi models belonging to different Thurston types such as $E^3$,
Nil (type $II$) and Sol ( type $VI_0$), because spaces belonging to
different Thurston types are not homeomorphic with each other.


\subsection{Self-similar Bianchi types}

It is of great interest to see if our results can be extended to
inhomogeneous and anisotropic universes. In general this is a very 
difficult mathematical problem. One simple class of inhomogeneous 
relatives of the Bianchi type universes is provided by their self-similar 
extensions, first found by Eardley (1974). It can be seen that there are 
no self-similar spacetimes with compact spaces if the self-similar group 
is not an isometry group and acts simply transitively. For, as Eardley 
shows, in such a case the self-similarity group, $H$, contains an isometry 
group, $G$, with ${\rm dim}G={\rm dim}H-1$, and each orbit of $G$ is 
two-dimensional. In the case of an open universe, this gives ${\mathbb 
R}^3$ the structure of a fibre bundle on a line ${\mathbb R}$ with fibres 
given by the orbits of $G$. If it can be compactified by some discrete 
group, $\Gamma $, then $\Gamma $ must be contained in the isometry group 
$G$. Hence taking the quotient with $\Gamma $may compactify the fibres but 
it leaves the underlying line ${\mathbb R}$ unchanged. Therefore the 
quotient space is at most a fibre bundle on a line with a compact 2-space, 
and cannot be a compact 3-space.

\section{Discussion}

We have shown that the stability of open and flat isotropic universes in the
space of spatially homogeneous initial data is strongly affected by the
global topology of the universe. When the topology is compact, all Class B
Bianchi type universes are forbidden unless they are isotropic. In
particular, this means that the universes of Bianchi type $VII_h$, and $V$
which contain the open Friedmann universes must be isotropic.

The most general anisotropic universes containing the flat Friedmann
universes are of Bianchi type $VII_0$ when the topology is non-compact.
However, in the presence of a compact topology the most general flat
anisotropic universes are of Bianchi type $I$ in the vacuum case and of type 
$VII_0$ in the presence of a perfect fluid. Thus, in the perfect fluid case
the stability properties of the isotropic models at late times isotropic
models is the same as for universes with non-compact topologies with
non-decay of anisotropic curvature distortions as $t\rightarrow \infty $ in
the radiation and dust dominated solutions, as found in earlier studies.

One consequence of this new behaviour in compact spaces is to change the
familiar conclusion that open universes are more general than flat universes
because they require more parameters for the specification of their initial
data. When homogeneous anisotropy is present the open universes are
forbidden while flat universes are allowed. This may have important
consequences for the assessment of the significance of the 'flatness'
problem in universes with compact topologies. It also indicates that in
compact open and flat universes there is a close link between the properties
of isotropy and flatness. These links should be investigated further in the
context of inhomogeneous cosmologies.

The local isotropy of  Bianchi $VII_h$ models with compact space implies
that the isotropy problem is replaced by the homogeneity problem for
inhomogeneous perturbations of isotropic universes. That is, if there are
some physical processes which homogenize a perturbed $VII_h$ compact
universe globally, they must automatically isotropize the universe because
homogeneous anisotropic $VII_h$ universes cannot exist. Alternatively, there
may be strong restrictions on the possibility of inhomogeneous open
universes with compact topologies. Our results also suggest the possibility
that the degree of anisotropy of the universe is constrained by the degree
of inhomogeneity if the universe is approximately described by a Friedmann
metric with a compact space of negative curvature locally. Such constraint
will apply even to inflationary universes and might be used to determine
whether or not the universe is negatively curved when the present universe
is almost flat due to inflation.

The strong impact of compactification upon the range of possible anisotropic
and homogeneous open universes is surprising. It shows one of the ways in
which topological restrictions can impose significant constraints on the
deviation of solutions of Einstein's equations from the simplest isotropic
cases that resemble the observed universe today. These results can be
generalized to other gravity theories which extend general relativity by
adding certain higher-order curvature terms to the lagrangian. These
higher-order terms are typically negligible at late times in ever-expanding
universes when the higher-order curvature scalars become smaller than the
linear terms contributed by general relativity.

One possible interpretation of our results is that the simultaneous presence
of anisotropy and spatial homogeneity is a very special combination. In the
cases of open or flat universes with non-compact topologies, the specialness
of this type of homogeneous anisotropy is disguised by the degrees of
freedom that are permitted for the homogeneous anisotropies. However, the
imposition of topological compactness is extremely restrictive and makes
some Bianchi geometries impossible in many anisotropic configurations. This
type of restriction was also evident when rotation was introduced into
closed compact Bianchi type $IX$ universes (Collins and Hawking 1973a,
Barrow et al 1985). A very strong limit on cosmic vorticity is created by
attempting to accommodate a complex rotational dynamics in a finite
positively curved space. The constraints on cosmic vorticity are accordingly
much weaker in non-compact flat and open universes.

Although the spatial compactness restricts anisotropy more weakly for
the Bianchi models including the spatially flat Friedmann universe,
some observable relations between anisotropy and inhomogeneity may
exist for particular space topologies. For example, if the topology of
the space is given by one of $T^3/{\mathbb Z}_2\times {\mathbb Z}_2$
and $T^3/{\mathbb Z}_k(k=3,4,6)$, then tilting of the fluid velocity
is forbidden if the universe is locally homogeneous. This suggests
that there may be a constraint on the degree of tilting by the degree
of inhomogeneity. Thus it will be interesting to investigate the
relation between anisotropy and inhomogeneity in the framework of
linear perturbation theory on spatially compact Bianchi models. There
will also be similar restrictions on certain types of anisotropic
stress, for example those contributed by magnetic and electric fields
or by collisionless massless particles, in these topologies.

It is a challenging and important problem to extend our analysis to the case
of inhomogeneous cosmologies. We found that this cannot be done for the
self-similar extensions of the Bianchi universes classified by Eardley
(1974). We know that inhomogeneous solutions of Einstein's equations that
are open or flat, with non-compact topologies, have initial data specified
on a spacelike hypersurface of constant time by 4 arbitrary functions of 3
spatial coordinates in the vacuum case and by 8 arbitrary functions in the
perfect fluid case. In the spatially homogeneous case these arbitrary
functions become arbitrary constants. While this suggests that the general
inhomogeneous solution may have parts that look locally like small
perturbations of the spatially homogeneous models, this need not be the case
for flat or open universes with compact topologies. Our study shows that the
topology of the universe can impose significant restrictions upon the type
of anisotropies that it can sustain.

\textbf{Acknowledgments}

We would like to thank Gary Gibbons, Janna Levin, and Stephen Siklos for
helpful discussions. JDB acknowledges support from the Royal Society, the
PPARC, and the University of New South Wales. HK was supported by the
Grant-In-Aid for the Scientific Research (C2) of the Ministry of Education,
Science, Sports and Culture in Japan (11640273).

\appendix

\section*{Appendix}

\begin{table}[tbp]
\begin{tabular}{llccccccc}
\hline\hline
\textbf{Symmetry} & \textbf{Space} & $Q$ & $P$ & $N_m$ & $N_f$ & $N$ & $N_s$
& $N_s$(vacuum) \\ \hline
&  &  &  &  &  &  &  &  \\ 
&  &  &  &  &  &  &  &  \\ 
II & ${\mathbb R}^3$ & 1 & 3 & 0 & 3 & 7 & 5 & (2) \\ 
& $T^3(n)$ & 4 & 4 & 0 & 3 & 11 & 9 & (6) \\ 
&  &  &  &  &  &  &  &  \\ 
II${\tilde\times}{\mathbb Z}_2$ & ${\mathbb R}^3$ & 1 & 3 & 0 & 2 & 6 & 4 & 
(2) \\ 
& $T^3(n)$ & 3 & 3 & 2 & 2 & 10 & 8 & (6) \\ 
& $K^3(n)$ & 3 & 3 & 0 & 2 & 8 & 6 & (4) \\ 
&  &  &  &  &  &  &  &  \\ 
II${\tilde\times} D_2$ & ${\mathbb R}^3$ & 1 & 3 & 0 & 1 & 5 & 3 & 2 \\ 
& $T^3(n)$ & 3 & 3 & 2 & 1 & 9 & 7 & 6 \\ 
& $K^3(n)$ & 3 & 3 & 0 & 1 & 7 & 5 & 4 \\ 
& $T^3(n)/{\mathbb Z}_2$ & 3 & 3 & 2 & 1 & 9 & 7 & 6 \\ 
& $T^3(n)/{\mathbb Z}_2\times{\mathbb Z}_2$ & 3 & 3 & 0 & 1 & 7 & 5 & 4 \\ 
&  &  &  &  &  &  &  &  \\ 
Isom(Nil) & ${\mathbb R}^3$ & 1 & 2 & 0 & 1 & 4 & 2 & 1 \\ 
& $T^3(n)$ & 2 & 2 & 2 & 1 & 7 & 5 & 4 \\ 
& $K^3(n)$ & 2 & 2 & 1 & 1 & 6 & 4 & 3 \\ 
& $T^3(n)/{\mathbb Z}_2$ & 2 & 2 & 2 & 1 & 7 & 5 & 4 \\ 
& $T^3(n)/{\mathbb Z}_2\times{\mathbb Z}_2$ & 2 & 2 & 1 & 1 & 6 & 4 & 3 \\ 
& $T^3(n)/{\mathbb Z}_k(k=3,4,6)$ & 2 & 2 & 0 & 1 & 5 & 3 & 2 \\ \hline\hline
&  &  &  &  &  &  &  & 
\end{tabular}
\caption{The number of dynamical degrees of freedom for locally homogeneous
systems with perfect fluid on the spaces of Thurston type Nil. }
\label{tbl:count:Nil}
\end{table}

\begin{table}[tbp]
\begin{tabular}{llccccccc}
\hline\hline
\textbf{Symmetry} & \textbf{Space} & $Q$ & $P$ & $N_m$ & $N_f$ & $N$ & $N_s$
& $N_s$(vacuum) \\ \hline
&  &  &  &  &  &  &  &  \\ 
&  &  &  &  &  &  &  &  \\ 
VI$_0$ & ${\mathbb R}^3$ & 2 & 3 & 0 & 4 & 9 & 7 & (3) \\ 
& $Sol(n)(n>2)$ & 3 & 3 & 0 & 4 & 10 & 8 & (4) \\ 
&  &  &  &  &  &  &  &  \\ 
VI$_0{\tilde\times}\{1,R_3(\pi)\}$ & ${\mathbb R}^3$ & 2 & 3 & 0 & 2 & 7 & 5
& (3) \\ 
& $Sol(n)$ & 3 & 3 & 0 & 2 & 8 & 6 & (4) \\ 
&  &  &  &  &  &  &  &  \\ 
VI$_0{\tilde\times}\{1,J\}$ & ${\mathbb R}^3$ & 2 & 3 & 0 & 2 & 7 & 5 & (3)
\\ 
& $Sol(n)(n>2)$ & 3 & 3 & 0 & 2 & 8 & 6 & (4) \\ 
&  &  &  &  &  &  &  &  \\ 
Isom$^+(Sol)$ & ${\mathbb R}^3$ & 2 & 3 & 0 & 1 & 6 & 4 & 3 \\ 
& $Sol(n)$ & 3 & 3 & 0 & 1 & 7 & 5 & 4 \\ \hline\hline
&  &  &  &  &  &  &  & 
\end{tabular}
\caption{The number of dynamical degrees of freedom for locally homogeneous
systems with perfect fluid on the spaces of Thurston type Sol. $J$ is a
discrete transformation which permutes two of the invariant basis.}
\label{tbl:count:Sol}
\end{table}

In this appendix we explain how to count the number of independent
parameters, $N_s$, specifying the diffeomorphism classes of solutions to the
Einstein equations for a locally homogeneous system.

First, we briefly review how solutions to the Einstein equations for a
locally homogeneous system are related to those for a homogeneous
system with a simply connected space. Let $M=\Sigma \times {\mathbb
R}$ be a locally homogeneous spacetime, and
$\tilde{M}=\tilde{\Sigma}\times {\mathbb R}$ be its universal
covering. The metric $g$ and the matter configuration $\Phi $ on $M$
can be lifted to $\tilde{M}$. The manifolds $M$ and $\tilde{M}$ are
related by $M=\tilde{M}/\Gamma $, where $\Gamma $ is a discrete group
of transformations on $\tilde{M}$ which preserve each constant time
slice, $\tilde{\Sigma}(t)=\tilde{\Sigma}\times \{t\}$. This lift
$(\tilde{g},\tilde{\Phi})$ must be invariant under the action of
$\Gamma $ and, from the assumption of local homogeneity, $\Gamma $ is
included in a larger group, $G, $ which acts simply homogeneously on
$\tilde{M}$.

In the synchronous coordinates on $\tilde{M}=\tilde{\Sigma}\times
{\mathbb R}\ni (x,t)$, the action of $G$ can be expressed as
time-independent transformations of $\tilde{\Sigma}$. Let $G^{\prime }
$ be a subgroup of $G$ which acts simply transitively on
$\tilde{\Sigma}$, and let $\chi ^0=dt$ and $\chi ^I$($I=1,2,3$) be the
invariant basis on $\tilde{\Sigma}$ with respect to $G^{\prime }
$. Then the lifted data on $\tilde{M}$, when expressed as components
with respect to the invariant basis, is specified by a set of
functions of time $X(t)$, which obey a set of first-order autonomous
ordinary differential equations with four constraints, corresponding
to the Hamiltonian constraint and the three momentum constraints. In
the vacuum case, $X(t)$ is given by the spatial metric components
$Q_{IJ}$ and their conjugate momenta $P^{IJ}$. In the perfect-fluid
case we must also include the fluid density $\rho $ and the three
spatial components of the fluid velocity, $u_I$. If $G$ is larger than
$G^{\prime }$, then $X$ must have additional symmetry. Since $\Gamma $
is included in $G$, these data are automatically invariant under
$\Gamma $. Hence, for each solution $X(t)$ to the Einstein equations
on $\tilde{M}$, the pair $(X(t),\Gamma )$ determines a solution to the
Einstein equations on $M$. The functions $X(t)$ are also uniquely
determined by the initial data $X(t_0)$ at some time $t=t_0$.

All the solutions on $M$ are obtained in this way. However, two
solutions on $M$ derived from $(X_1(t_0),\Gamma _1)$ and
$(X_2(t_0),\Gamma _2)$ may be connected by a diffeomorphism on $M$,
and so be physically equivalent. It can be shown that this happens
when and only when the two solutions are related by a time
translation, or the two initial data are connected as
$X_2(t_0)=f_{*}X_1(t_0)$ and $\Gamma _2=f\Gamma _1f^{-1}$ by a so
called
\textit{homogeneity-group-preserving diffeomorphism} (HPD) $f$ which
preserves the symmetry group $G$ in the sense that $fGf^{-1}=G$. All the
HPDs form a group, $\mathrm{HPDG}(G)$, which is the normalizer group of $G$
in $\mathrm{Diff}(\tilde{M})$ in mathematical terminology.

This gauge freedom can be removed to produce a unique specification of
each diffeomorphism class in the following way. First, we introduce
the moduli parameters as a coordinate system of a maximal submanifold
$\mathcal{M}$ in the space of $\Gamma $ which is transversal to orbits
of the action of $\mathrm{HPDG}(G)$. If the isotropy group $H$ of the
action of $\mathrm{HPDG}(G)$ is trivial, the gauge freedom is
completely removed by this procedure.  On the other hand, if the
isotropy group is non-trivial, we further introduce a set of
parameters as a coordinate system on a submanifold in the space of
$Q_{IJ}$ which is transversal to orbits of the $H$ action. If $H$
still has a non-trivial isotropy group, we further apply the same
procedure to other variables, say $P^{IJ}$. Eventually, we obtain a
set of parameters specifying the diffeomorphism classes of initial
data for the locally homogeneous system, i.e., a coordinate system for
the phase space $\Gamma _{\mathrm{inv}}(M,G)$. The number of
parameters classifying the diffeomorphism classes of solutions is
obtained by subtracting $1$ from the dimension of $\Gamma _
{\mathrm{inv}}(M,G)$.

This method can be also applied to the standard spatially homogeneous
system with a simply connected space by considering the case in which
$\Gamma $ is trivial. For example, in the vacuum Bianchi $I$ system,
$X$ is given by $(Q_{IJ},P^{IJ})$, and $\mathrm{HPDG}({\mathbb R}^3)$
is given by $IGL(3,{\mathbb R})$ whose element induces a similar
transformation of the matrices $Q$ and $P$. By $\mathrm{HPDG}$ $Q$ can
be set to the unit matrix. The isotropy group acts on $P$ as
$O(3)$. Hence $P$ can be put into a diagonal form. After this
reduction, the action of the residual $\mathrm{HPDG}$ on the initial
data becomes trivial. Hence, by taking account of the Hamiltonian
constraint and the time translation freedom, we find that the
equivalence class of the solutions are specified by a single
parameter, which will correspond to one of the Kasner indices.

As a procedure to determine $N_s$ for non-compact Bianchi models, our
method is more complicated than those used by Siklos (1978) or by
Ellis and MacCallum (1969). In their methods $\Gamma _{\mathrm{inv}}$
is simply obtained as $S/GL(3)$ or $S/SO(3)$, where $S$ is the space
of components with respect to a general invariant basis and its
structure constant, $(Q_{IJ},P^{IJ},C^I{}_{JK})$, and that with
respect to an orthonormal invariant basis, $(P^{IJ},C^I{}_{JK})$,
respectively. In contrast, in our method, we need information on HPDG,
which is in general obtained by a long calculation. However, in
calculating $N_s$ for compact locally homogeneous system, this
information is indispensable to determine the moduli degrees of
freedom correctly.

Tables \ref{tbl:count:E3}--\ref{tbl:count:Sol} summarize the number of
dynamical degrees of freedom obtained by this procedure for locally
homogeneous systems with perfect fluid on compact spaces with Thurston
type $E^3$, Nil and Sol as well as the corresponding ones for Bianchi
types $I,VII_0,II,$ and $VI_0$ on ${\mathbb R}^3$. In these tables,
$Q,P, N_m$ and $N_f$ denote the number of independent variables to
specify the space metric, the extrinsic curvature, the moduli freedom
and the fluid freedom respectively, in the diffeomorphism invariant
phase space $\Gamma_{\mathrm{\ inv}}(M,\tilde G)$ with an invariance
group of the data on a space type $M$. $N$ and $N_s$ represent the
dimensions of $\Gamma_{\mathrm{inv}}(M,\tilde G)$ and of the solution
space. For comparison, $N_s$ for vacuum systems are given on the final
column where $(*)$ implies that the corresponding vacuum system has a
higher discrete symmetry. Note that the total number of dynamical
degrees of freedom in this table does not take the Hamiltonian
constraint into account, and hence is greater by $1$ than the
dimension of $\Gamma _{\mathrm{inv}}(M,G)$ defined above. It should be
also noted that in the case of Bianchi type $I$, the lowest symmetry
is not ${\mathbb R}^3$ and there are additional discrete symmetries
$D_2=\{1,R_1(\pi ),R_2(\pi ),R_3(\pi )\}$ corresponding to rotations
by angle $\pi $ around each coordinate axis because the fluid velocity
is obliged to be orthogonal to constant time hypersurfaces due to the
momentum constraints. If this $D_2$ symmetry exists, the spacetime
metric can be put into diagonal form with respect to the
time-independent invariant basis. In contrast, if the symmetry group
$G$ does not contain $D_2$, such time-independent diagonalization is
not possible even if $Q$ and $P$ have three dynamical degrees of
freedom respectively. In such cases, the diagonalization of the metric
requires time-dependent HPDs, which produce a non-vanishing shift
vector.

If one allows for the time-dependent HPDs, one can always put the
variables $X$ into the form for the system on ${\mathbb R}^3$ with the
same symmetry.  This transformation transfers some degrees of freedom
in $Q$ and $P$ back into the moduli freedom and makes the moduli
parameters time-dependent. This enlarged moduli freedom is often used
as the definition of moduli freedom in the literature. However, the
description of dynamics becomes more complicated using this approach.


\end{document}